\documentclass[12pt]{iopart}
\usepackage{amssymb}
\usepackage{booktabs}
\usepackage{multirow}
\usepackage{xcolor}
\usepackage{lineno}
\usepackage{dcolumn}
\usepackage{url}
\usepackage{hyperref}
\usepackage{graphicx}

\begin{document}
\interfootnotelinepenalty=10000

\title[]{Automated visual inspection of CMS HGCAL silicon sensor surface using an
ensemble of a deep convolutional autoencoder and classifier}

\author{Sonja Gr\"{o}nroos\textsuperscript{1}, Maurizio Pierini\textsuperscript{1}, Nadezda Chernyavskaya\textsuperscript{1}}
\address{$^1$CERN, 1211 Geneva 23, Switzerland}
\ead{sonja.gronroos@cern.ch}

\begin{abstract}
More than a thousand 8" silicon sensors will be visually inspected to look for anomalies on their surface during the quality control preceding assembly into the High-Granularity Calorimeter for the CMS experiment at CERN.
A deep learning-based algorithm that pre-selects potentially anomalous images of the sensor surface in real time has been developed to automate the visual inspection.
The anomaly detection is done by an ensemble of independent deep convolutional neural networks: an autoencoder and a classifier.
The performance is evaluated on images acquired in production.
The pre-selection reduces the number of images requiring human inspection by 85\%, with recall of 97\%.
Data gathered in production can be used for continuous learning to improve the accuracy incrementally.
\end{abstract}

\vspace{2pc}
\noindent{\it Keywords}: Anomaly detection, autoencoder, convolutional deep neural networks, silicon sensors, quality control, visual inspection
%\maketitle

\section{Introduction}

\begin{figure}[h!]
\begin{centering}
  \includegraphics[width=8.5cm]{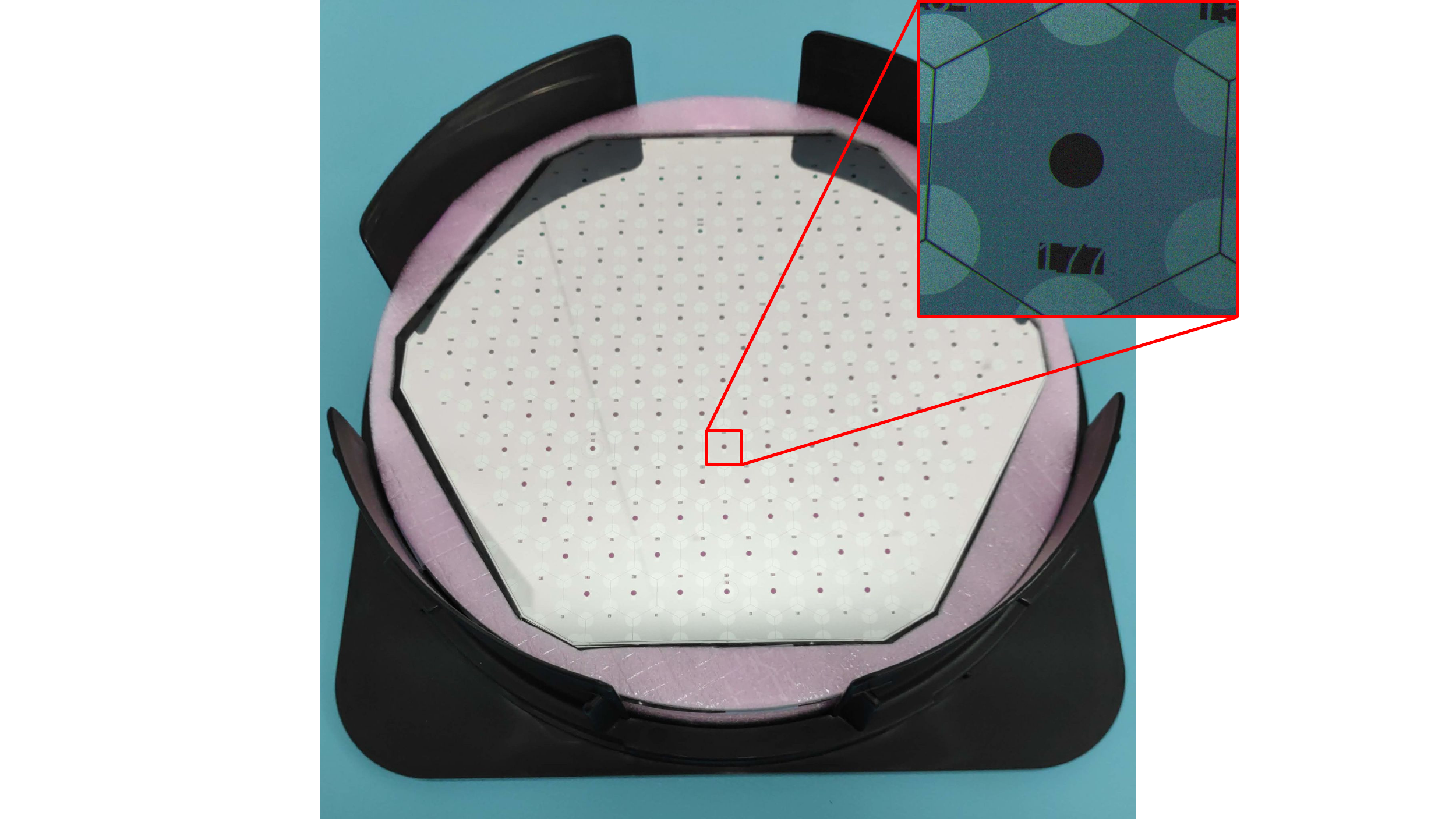}
  \includegraphics[height=4.5cm]{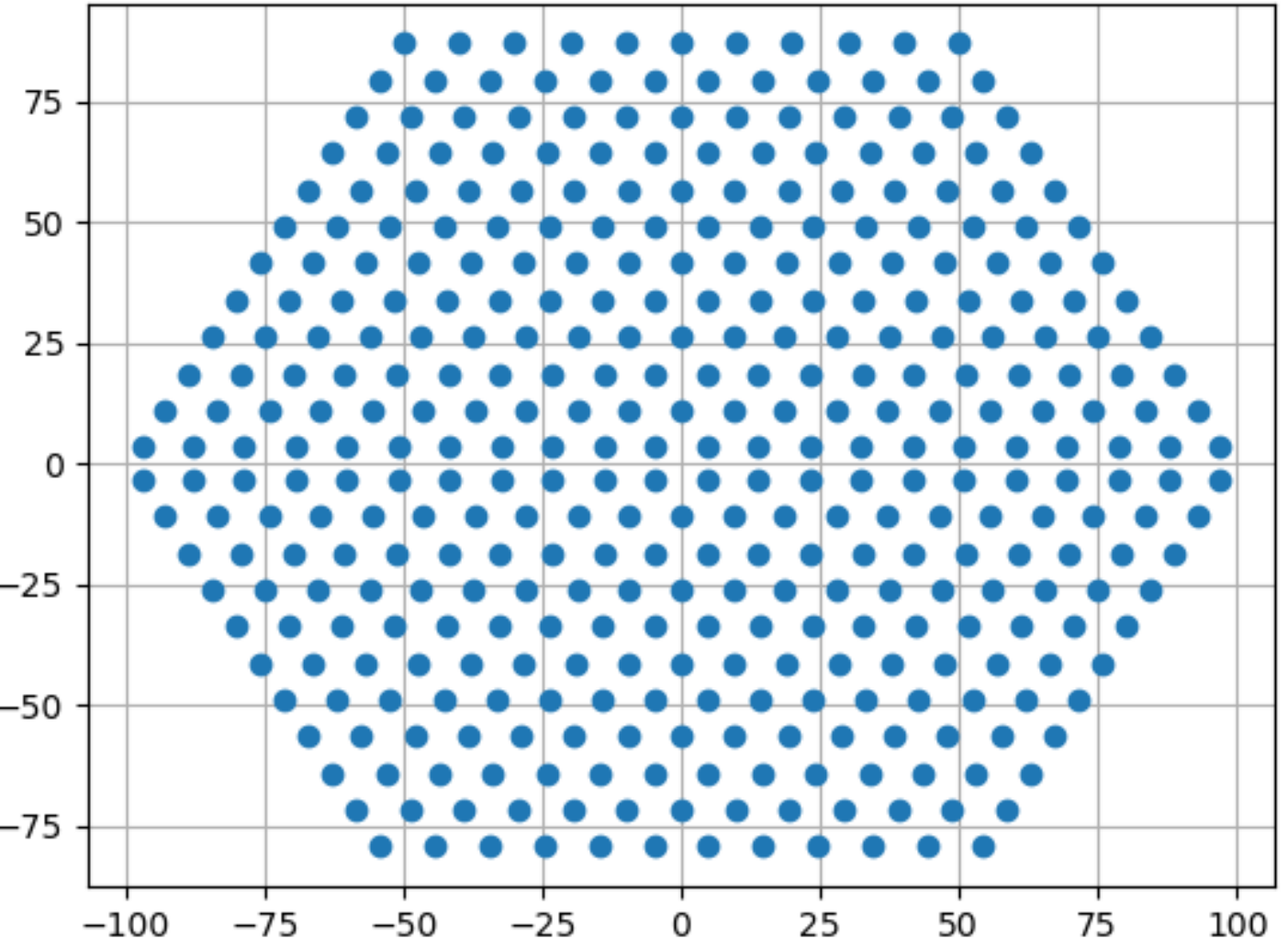}
\caption{(left) An HGCAL silicon sensor wafer, with a pad zoomed in. The diameter of the sensor is 8". (right) An example of a scan map.}
\label{fig:sensor_map}
\end{centering}
\end{figure}

Silicon sensors are used in high-energy physics experiments due to their sufficient radiation tolerance, energy resolution and cost-effectiveness.
In the high radiation area, the active element of the High-Granularity Calorimeter (HGCAL) \cite{CERN-LHCC-2017-023}, which will replace the endcap calorimeters of the CMS \cite{CMS} experiment at the Large Hadron Collider (LHC) \cite{LHC}, will consist of more than 27,000 hexagonal 8" silicon sensor wafers to achieve unprecedented transverse and longitudinal segmentation. 
An HGCAL sensor is shown in figure \ref{fig:sensor_map} (left).
The producer of the sensors is Hamamatsu Photonics K.K, and the sensors will arrive to CERN in batches.
In order to ensure that the sensors meet the criteria for operation at the LHC, a fraction (5\%) of each batch will undergo quality control (QC) in a dedicated clean room at CERN.
Thus, more than a thousand sensors will be processed over the course of several months.

\begin{figure}[h!]
\centering
  \includegraphics[width=7cm]{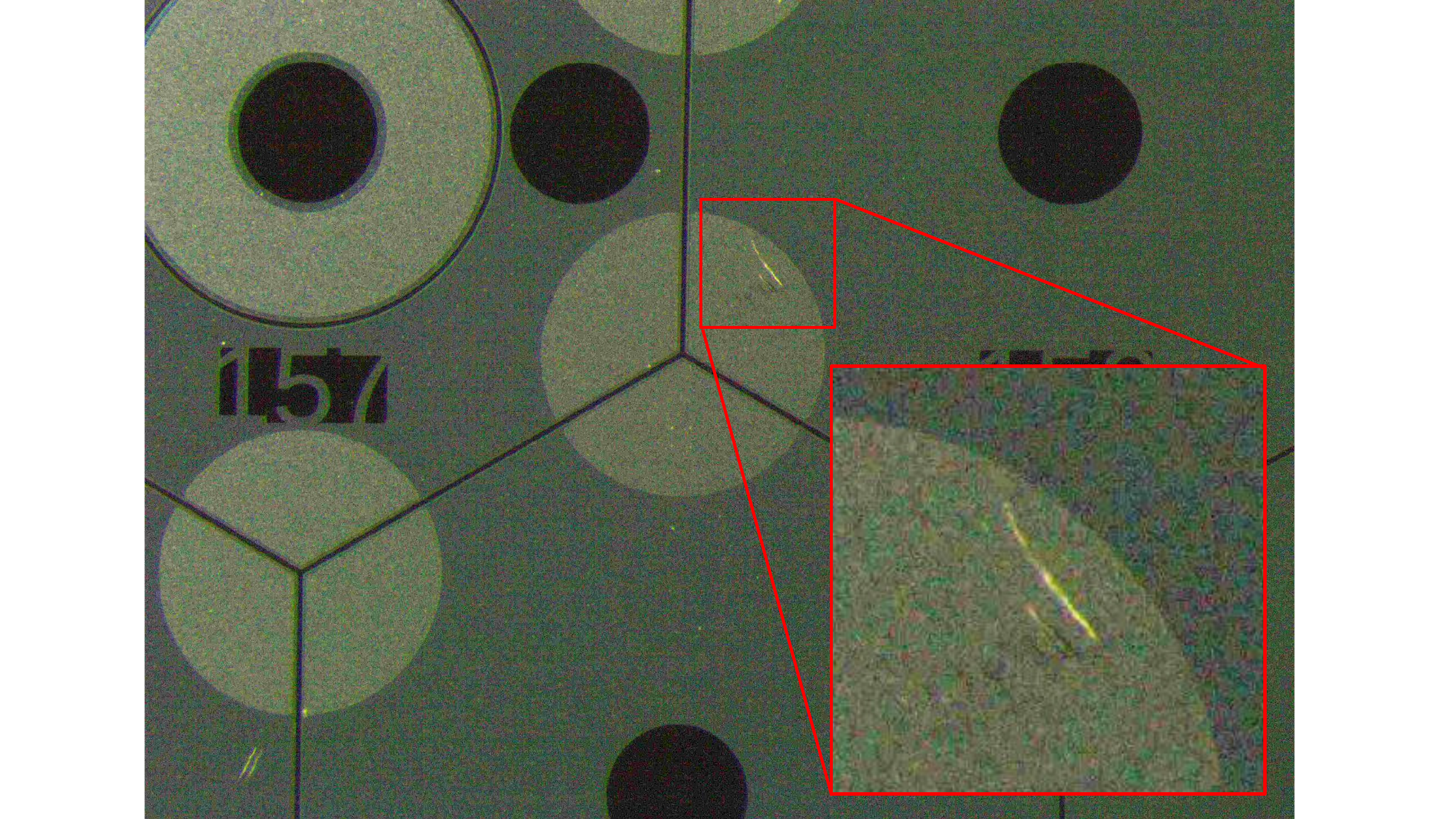}
  \includegraphics[width=7cm]{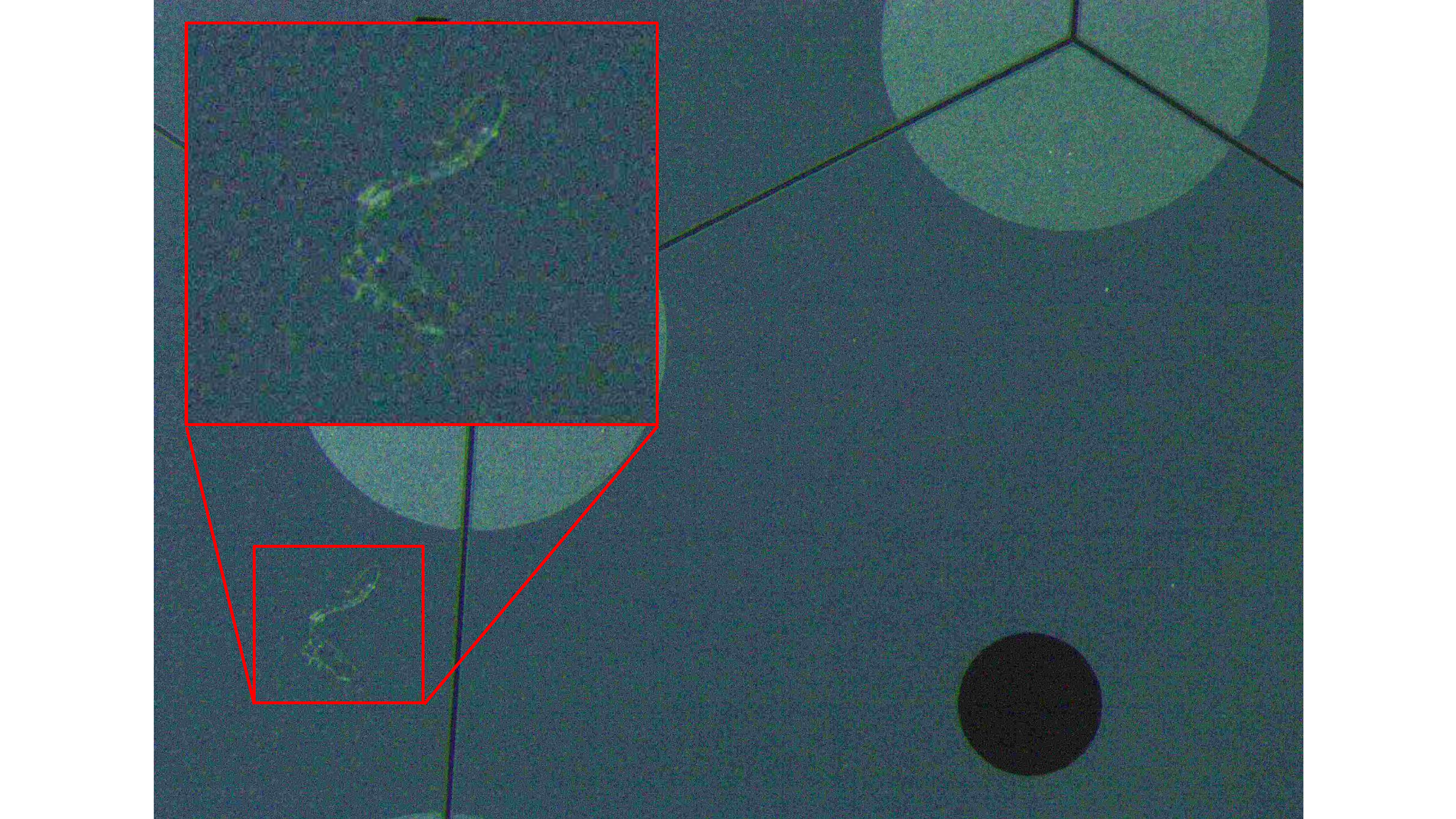}
\caption{Scan images in RGB format for anomalous sensors with a scratch (left), and a dust particle (right).
The difference in color is induced by lighting conditions.}
\label{fig:dust_scract}
\end{figure}

The QC procedure adopted during the construction of the CMS silicon trackers involved a visual inspection (VI) in addition to quantification of several electrical properties \cite{CMS_QC}.
Similarly, a major part of the QC of the HGCAL sensors is the electrical characterization of the sensors, during which the sensors are biased up to 1000 V \cite{elec_char}.
Defects and dust on a sensor surface can potentially lead to an electrical failure of the sensor.
Examples of a typical defect, a scratch, and a dust particle are shown in figure \ref{fig:dust_scract}.
Given that these defects are rare and unwanted, they are referred to as anomalies.
The anomalies can occur during manufacturing, packaging, delivery, or associated handling of the sensors. In an effort to prevent failures, the sensor surface is visually inspected and cleaned prior to the electrical characterization.

Aside from exceptionally severe scratches, most of the anomalies are invisible to the naked eye.
Traditionally, silicon VI is carried on manually with the help of a microscope.
However, dozens of square meters of sensor surface will be inspected during the assembling of HGCAL, and therefore, a standardized and automated method must be in place.
Previous to this work, hundreds of microscope images were taken using a scan program, and the images were inspected on a computer monitor by a human operator.
This work presents a deep learning-based pre-selection algorithm (PSA) that fully automates the VI. In addition, the PSA is believed to reduce human bias in the VI. The PSA is built upon the proof-of-concept work described in \cite{goals}.
Although the PSA is presented in the context of HGCAL sensor QC, the same approach could be applied to similar use cases of automating the VI of images.

The PSA proposed in this work detects anomalous scan images via an ensemble of a deep convolutional autoencoder (AE) and a deep convolutional classifier neural network. 
The AE acts on each scan image, and the classifier acts on patches of the image, allowing for localization of anomalous areas (annotation).
The PSA has been deployed in a clean room at CERN, and data gathered in production are used to evaluate the performance of the anomaly detection.
The performance is mainly measured using two metrics: First, the False Negative Rate (FNR), defined as
\begin{equation}
\label{eq:fnr}
    \textrm{FNR} = \frac{\textrm{False Negative}}{\textrm{False Negative} + \textrm{True Positive}}
\end{equation}
should be minimized for a reliable PSA.
Second, a relatively large False Positive Rate (FPR), defined as
\begin{equation}
\label{eq:fpr}
    \textrm{FPR} = \frac{\textrm{False Positive}}{\textrm{False Positive} + \textrm{True Negative}}
\end{equation}
is allowed, but an upper limit of 10\% is set to sufficiently automate the VI.

This paper is structured as follows. The data acquisition and characteristics are described in section \ref{sec:setup and data set}. An introduction to automated VI is given in section \ref{sec:overview}, followed by a description of the proposed architecture and the model training process in section \ref{sec:approach}. The results of the deployment at CERN are presented in section \ref{sec:results}. In section \ref{sec:discussion}, the efficiency and causes of incorrect predictions are discussed, and the proposal for continuous improvement of the anomaly detection capabilities is presented. Finally, conclusions are provided in section \ref{sec:conclusion}.

\section{Setup and data set}
\label{sec:setup and data set}

A custom semi-automated VI system has been implemented in the clean room for HGCAL sensor testing. 
Using a programmable xy-stage, the sensor is moved beneath a combination of a microscope and a camera in a scan pattern. An example of a scan map is shown in figure \ref{fig:sensor_map} (right), where 385 images are taken.
A scan image, referred to as a \emph{whole image}, contains 2720 $\times$ 3680 pixels and is stored in Bayer format \cite{bayer}. 
The Bayer format is a particular arrangement of RGB color filters common to camera systems, which retains the color information but reduces the required bits per pixel from 24 to 8 bits. 
Examples of whole images in RGB format are shown in figure \ref{fig:dust_scract}.

The images acquired during the semi-automated VI require human inspection. 
A small fraction of the images of a typical scan are anomalous, meaning that the operator has to inspect hundreds of normal images to find the anomalous ones.
This makes the semi-automated VI tiring, slow, and typically not 100\% effective due to visual fatigue. Moreover, it can be biased by inspector's overexposure to normal images which are prevalent in the data set. In addition, since multiple inspectors with varying experience and alertness share the QC task, the VI is further biased by their subjectivity.

The environmental conditions, such as zoom level, sensor alignment underneath the microscope and lighting conditions, can change in between scans, and the PSA must be invariant to these changes.
An example of changing lighting conditions between the measurements is demonstrated in figure \ref{fig:dust_scract}, where the left and right images differ in the overall hue.
As the PSA is integrated into the data acquisition of the semi-automated system, it must be real-time.

Taking into consideration the data imbalance, variable environment, and requirements for accuracy and speed, the PSA presented in this paper was developed using images acquired during semi-automated VI of 50 sensors.
The data were acquired in batches over the course of several months, and consists of more than 25,000 images.
Fifteen sensor scans acquired after the deployment of the PSA are used to evaluate its performance.

\section{Overview of existing methods for automated visual inspection}
\label{sec:overview}

The task of the PSA is identification and localization of rarely occurring outliers in data.
Sometimes, an analytical approach in the form of a series of image processing filters and functions can be used to detect anomalies instead of more complex methods such as deep learning.
For example, anomalies have been detected from images of the silicon strip sensors of the Inner Tracker of the ATLAS detector~\cite{Atlas} using methods such as a Gaussian filter and Sobel derivatives \cite{kanopoulos1988design,openCV_processing}. 
However, due to changing environmental conditions (including room lighting) and characteristics of the normal HGCAL sensor surface, these methods cannot produce robust results. Thus, deep learning, and specifically deep convolutional neural networks (CNNs) \cite{CNN_general1}, are explored in this work.
CNNs are known to perform well in image classification tasks. Several classifier networks have been developed, such as the VGG16 \cite{VGG16}. Characteristic to its architecture are sequential convolutional layers with small 3 $\times$ 3 filters and 2 $\times$ 2 max pooling layers. However, classifier networks are not object detectors, as they do not indicate the location of the object. Instead, a widely used network for object detection is the Region-based Convolutional Neural Network (R-CNN) \cite{R_CNN}, which performs object detection via three distinct networks. 
The first network is a region proposal network, which extracts up to 2,000 regions of interest from the input image. 
The regions of interest are passed onto the second model, which is a CNN that extracts the features of each region.
Finally, a classifier CNN is applied on the features to produce the classification output in the form of bounding boxes.
The R-CNN has been used in automating the VI of silicon micro-strip sensors \cite{lavrik}.

Unfortunately, the R-CNN is too slow for real-time object detection, as thousands of iterations are required per image to produce the detection output. Thus, faster versions of the model, such as the Faster R-CNN \cite{faster_R_CNN}, have been developed. 
However, a preferred approach for real-time object detection is to perform both the region and feature extraction in the same network and with a single iteration for an image. An example of such a network is the You Only Look Once (YOLO) network \cite{YOLO}, which splits an image into cells via a grid and predicts $n$ bounding boxes and the class probabilities for each cell. In the context of the HGCAL project, the use of VGG16 and a version of YOLO known as YOLOv4-tiny have been studied for the VI of wire bonds of the sensor modules \cite{CNN_bonding}.

The above discussed CNNs have to be trained in a supervised fashion, with training samples from all classes. 
Self-supervised anomaly detection can be implemented using AEs, which are composed of two neural networks. 
The first network, known as the encoder, reduces the dimensionality of the input data into a representation referred to as the latent space. 
The second network is a decoder, which reconstructs the latent space back into the original dimensionality of the input. 
An AE is trained by minimizing the loss function, expressed as the reconstruction error.
The reconstruction error is usually quantified as the mean absolute error between the input $\textbf{y}$ and the reconstructed output $\hat{\textbf{y}}$, and defined as 
\begin{equation}
\label{eq:L1}
    \textrm{L1} = \frac{1}{N}|\hat{\textbf{y}}-\textbf{y}|.
\end{equation}
The squared error, defined as
\begin{equation}
\label{eq:L2}
    \textrm{L2} = \frac{1}{N}(\hat{\textbf{y}}-\textbf{y})^{2}
\end{equation}
can also be used. 
For anomaly detection, the AE is trained on only normal samples, and thus, it will reconstruct anomalies poorly.
Convolutional AEs have been used for anomaly detection from image-like data in multiple applications \cite{CAE_ex1, CAE_ex2, AE3}, also at the LHC \cite{AE_LHC1, AE_LHC2}. 
For images, the reconstruction error is calculated pixel-wise, and localized increases in the reconstruction error indicate anomalies.

\section{Proposed approach}
\label{sec:approach}

\begin{figure}[ht]
\centering
\includegraphics[width=13cm]{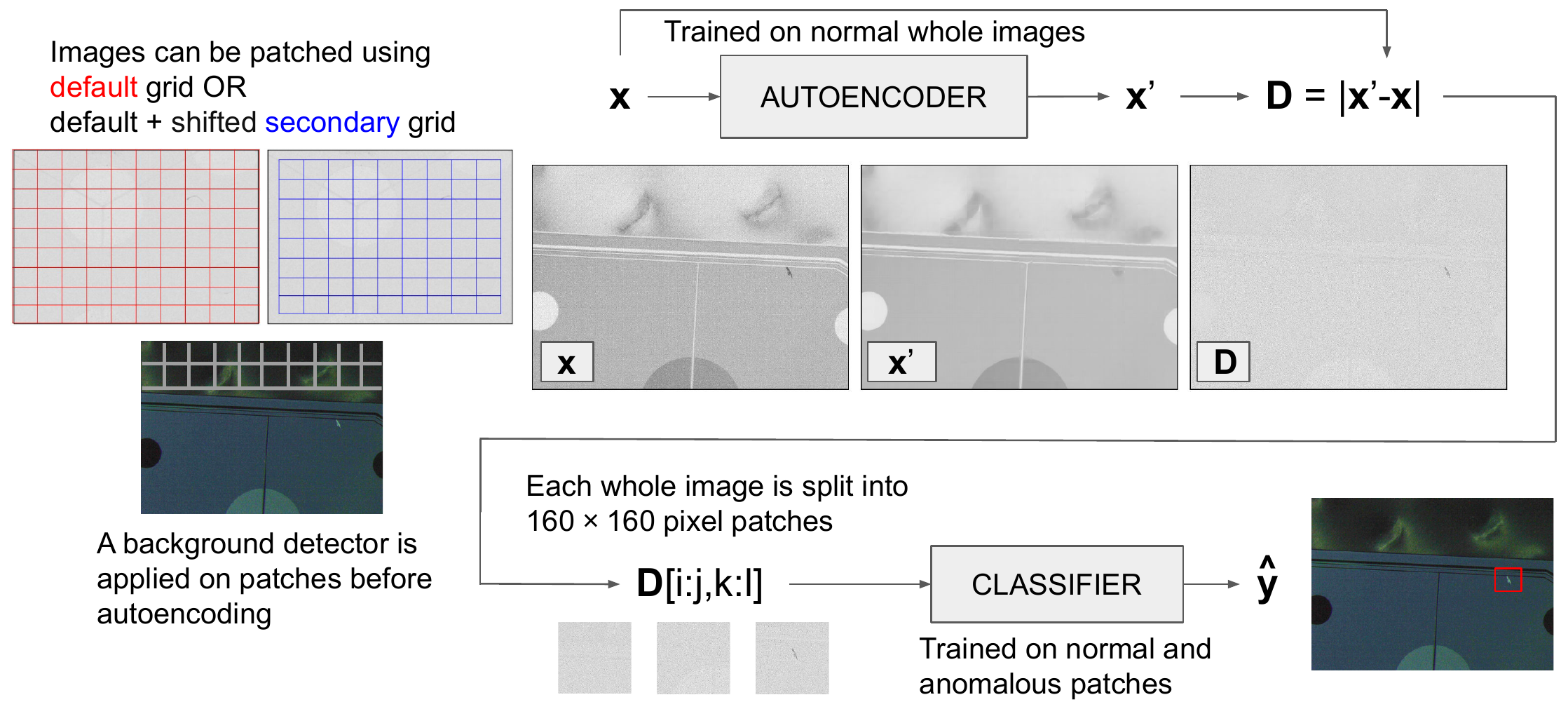}
\caption{The inference pipeline of the anomaly detection. The input image $\textbf{x}$ is processed with an autoencoder, and the pixel-wise reconstruction error $\textbf{D}$ is calculated between the input and the autoencoder output $\textbf{x}'$. $\textbf{D}$ is given as input to the anomaly detecting classifier in patches.}
\label{fig:inference_pipe}
\end{figure}

In this section, the architecture of the PSA is described.
The full \emph{inference pipeline} of a whole image consists of the following steps: 
\begin{enumerate}
\item Apply a patching grid.
\item Apply a background detecting classifier, referred to as the \emph{background detector}, to patches of the whole image.
\item Apply an AE to the whole image and calculate the reconstruction error as the pixel-wise absolute difference $\textbf{D}$.
\item Apply an anomaly detecting classifier, referred to as the \emph{anomaly detector}, to the patches of $\textbf{D}$.
\end{enumerate}
Thus, each whole image is iterated over three times, and notably, the images are kept in the Bayer format throughout the entire inference pipeline.
The anomaly detection process is schematically shown in figure \ref{fig:inference_pipe}. 
The models were implemented using Keras \cite{keras} and TensorFlow 2 \cite{tf}, and trained on a NVIDIA GeForce GTX 1080 GPU \cite{cuda}. The hyperparameters of all three models were optimized manually using their respective losses as a metric.

\subsection{Patching}

\begin{figure}[h!]
\centering
  \includegraphics[width=12cm]{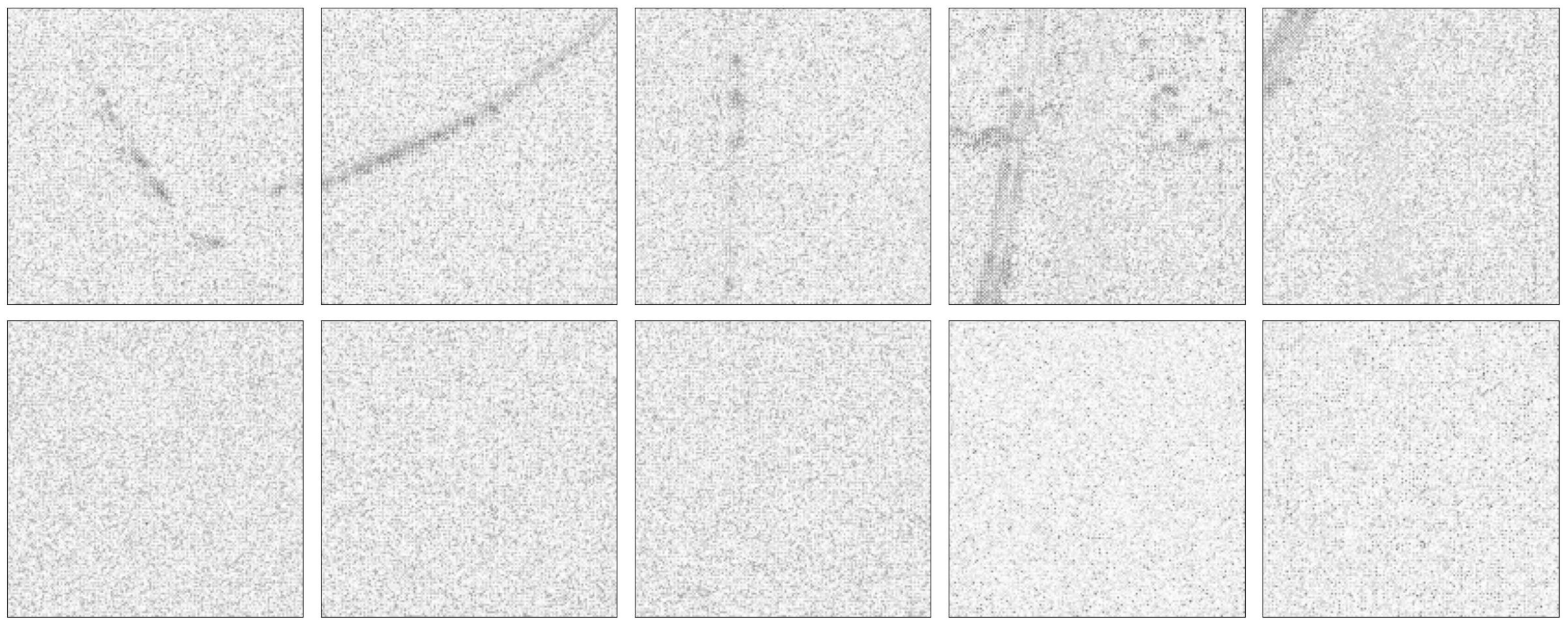}
  \caption{Sample of anomalous (top) and normal (bottom) patches of the sensor surface used to train the anomaly detector.}
\label{fig:patches_ex}
\end{figure}

The whole images are split into patches using a fixed grid.
The patches are 160 $\times$ 160 pixels in size, and the default grid covers the entire image, resulting in 17 $\times$ 24 patches. 
The patching can be considered as a simplified version of the region proposal for R-CNN.
For training the background and anomaly detectors, the patches are given the corresponding binary labels: 0 for sensor surface/normal and 1 for background/anomalous.
Examples of anomalous and normal patches processed with the AE are shown in figure \ref{fig:patches_ex}.

The main reasons for patching are that the fraction of area covered by an anomaly is much larger for an anomalous patch than for an anomalous whole image, and that the input size for a classifier becomes smaller. Also, the patching allows the general location of the anomaly in the whole image to be computed. In addition, data augmentation can be done more efficiently by applying it to anomalous patches only, and a class can be under-sampled flexibly from patched data.

\subsection{Background elimination}

During a typical scan, approximately 15\% of the images contain the background, referring to the surface the sensor lies on. 
Occasionally, the background, which is typically a black sheet of plastic, contains features which can be incorrectly selected as anomalies.
While FNR is the most important metric to optimize, to further reduce the FPR of the PSA, a background detector is applied to the patches before the whole image is autoencoded.
Anomalies in the background patches are ignored.
The background detector was built as standalone to allow its elimination from the inference pipeline.

A CNN with four convolutional layers using the ReLU activation, followed by dropout layers with a rate of 20\%, and totaling 84,401 trainable parameters, was trained on images sampled from the training data. 
The patches were given binary labels corresponding to sensor surface and background. 
The training data and parameters are described in table \ref{tab:CNN_training_bg_AE}. 
Classification between the sensor surface and background is a trivial task for a deep CNN, and a test accuracy of over 99\% was achieved.

\begin{table}[ht]
\centering
\caption{Summary of training data and parameters for the background detector and the autoencoder. For the background detector, Class 0 refers to sensor surface and Class 1 to background.}
\label{tab:CNN_training_bg_AE}
\vspace{10pt}
\begin{tabular}{@{}llllll@{}llllll@{}}
\toprule
                 &   Background detector  &  Autoencoder \\ \midrule
Whole images                 &   962  &  16,000  \\ 
Class 0 training patches      &  347,833 &  -  \\ 
Class 1 training patches      &  20,183  &  -  \\
Batch size     & 256  & 1   \\ 
Epochs         & 55   & 277  \\ 
Optimizer      & $Adam$  &$Adam$ \\
Learning rate  & 10$^{-4}$    & 10$^{-4}$ \\ 
Loss           & Binary cross-entropy  & L2\\ \bottomrule
\end{tabular}
\end{table}

\subsection{Anomaly enhancement using an autoencoder}

\begin{figure}[h!]
\centering
  \includegraphics[width=12cm]{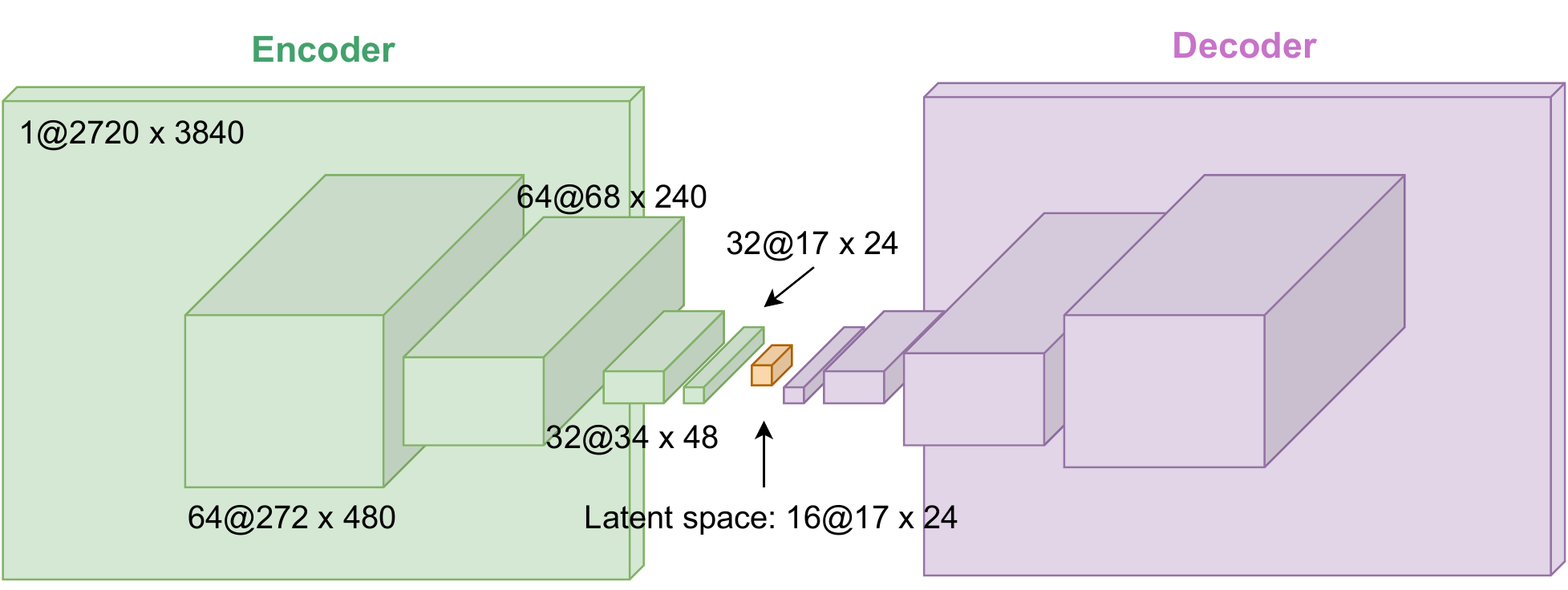}
  \caption{An illustration of the dimensions of the autoencoder, which consists of two convolutional neural networks: the encoder and the decoder. The autoencoder takes a whole image as an input, and the compression factor into the latent space is 1,600.}
  \label{fig:AE_arch}
\end{figure}

The structure of the AE is schematically shown in figure \ref{fig:AE_arch}.
The encoder consists of five convolutional layers, and
in mirror-like fashion, the decoder consists of five transposed convolutional layers.
The compression factor of the AE is 1,600.
The Exponential Linear Unit \cite{ELU} was used as the activation function.
For the training, the L2 loss and \emph{Adam} \cite{Adam} optimizer with a learning rate 10$^{-4}$ were used.
In total, the AE has 126,353 trainable parameters, and it was trained with 16,000 normal whole images for 277 epochs, until validation loss reached a plateau. 
Due to memory constraints, the batch size had to be set to one. The training data and parameters are summarized in table \ref{tab:CNN_training_bg_AE}.  

\begin{figure}[h!]
\centering
  \includegraphics[width=13cm]{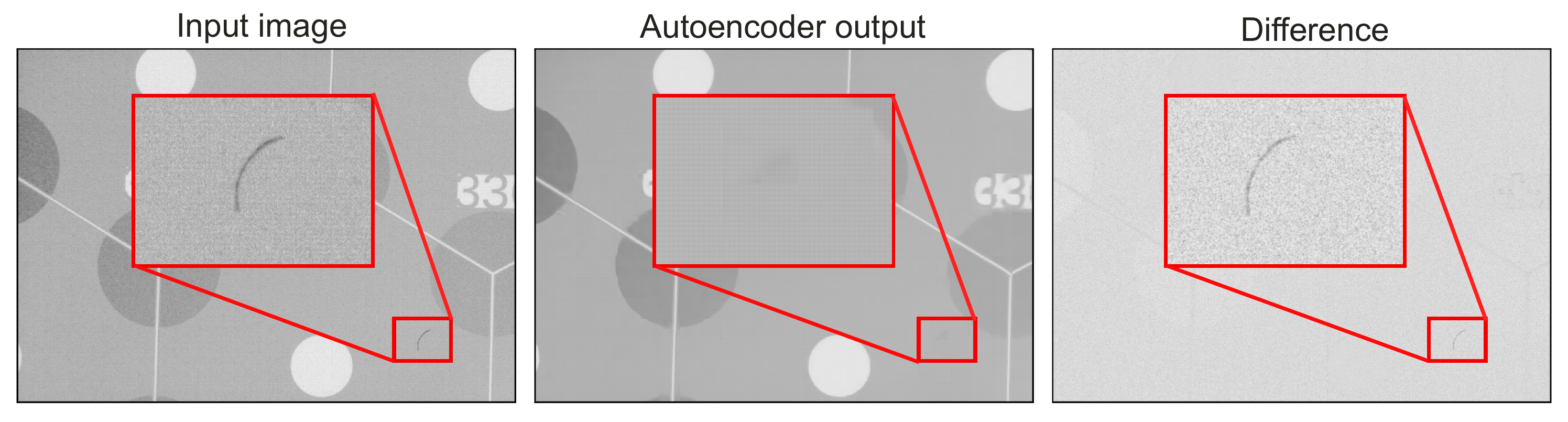}
   \caption{An example of the autoencoder output (center) for an anomalous input image (left). (right) Reconstruction error, measured as the absolute pixel-wise difference between the autoencoder output and the input. A dust particle is enhanced compared to the normal area, and can be easily isolated. The anomaly is zoomed in on all images.}
  \label{fig:AE_dust}
\end{figure}

The AE can be interpreted as a data pre-processing step that makes the subsequent anomaly detection and its localization more robust against environmental changes. 
The normal and constant features in the images are reduced, while the anomalies are enhanced. 
An example of how the AE reconstructs an anomalous whole image is shown in figure \ref{fig:AE_dust}.

\subsection{Anomaly detection}

\begin{figure}[h!]
\centering
  \includegraphics[width=7cm]{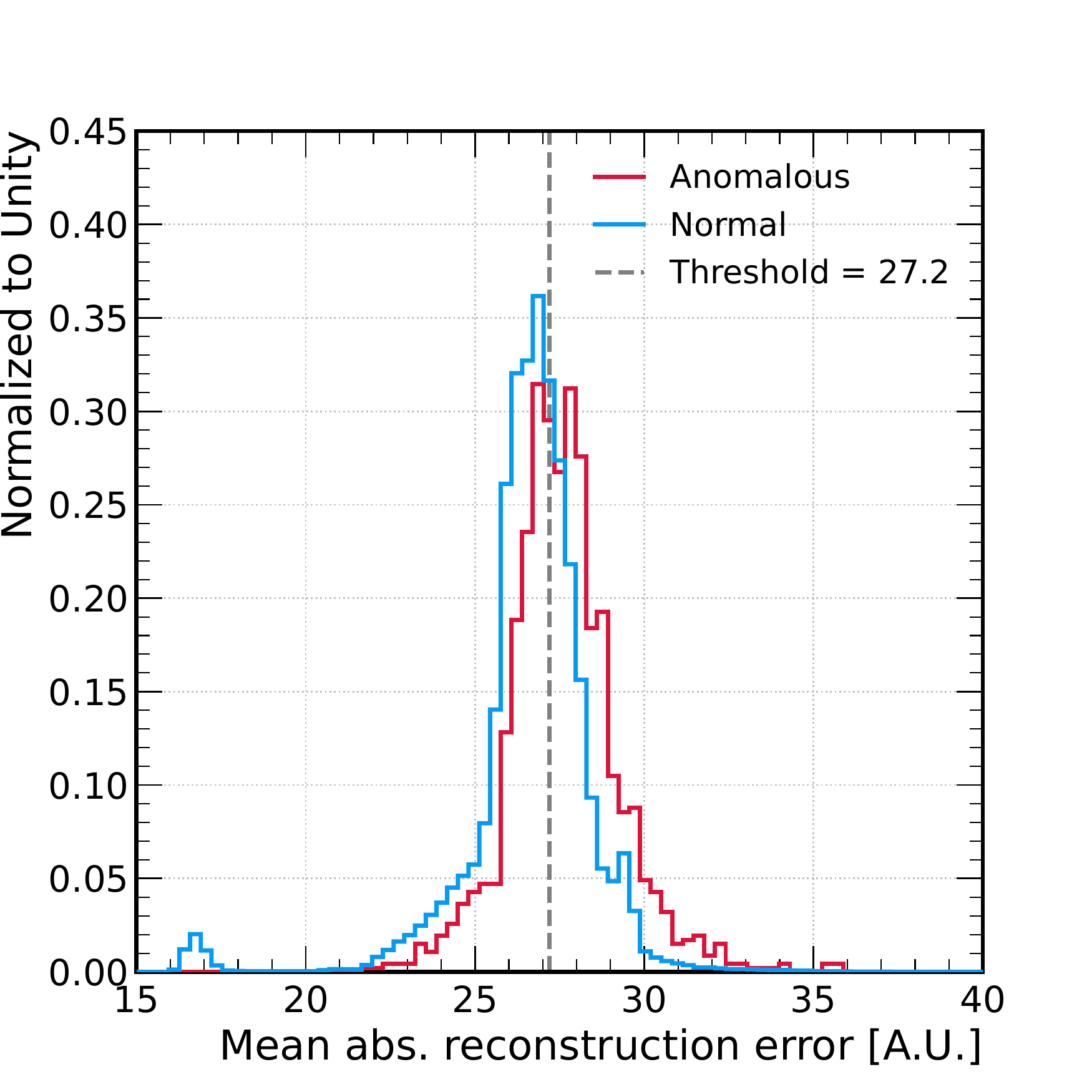}
   \caption{Normalized autoencoder reconstruction error for anomalous and normal patches. A threshold to classify new samples is also illustrated.}
  \label{fig:AE_hist}
\end{figure}

First, the performance of the AE as a standalone anomaly detector was studied using 1,465 anomalous and 225,370 normal patches as training data.
A threshold for the reconstruction error was determined based on a validation data set, consisting of 157 anomalous and 27,179 normal patches, such that the validation FNR and FPR were minimized.
An increase in AE reconstruction error for anomalous patches is visible in figure \ref{fig:AE_hist}, where the selected threshold is also indicated\footnote{Some patches, e.g. of the black areas on the sensor surface, are easy to reconstruct by the AE, appearing as the small bump at the lower range of the reconstruction error in figure \ref{fig:AE_hist}.}. 
However, as the distributions overlap greatly, the AE reconstruction error cannot be used as a robust enough classifier. The test FNR is 27\%, while FPR is 37\%. Therefore, while the AE works very well as a pre-processing step that enhances the anomalies, it cannot be efficiently used to detect and localize them. To tackle anomaly detection and localization, an additional classifier was trained to detect the anomalies. 

\begin{figure}[h!]
\centering
  \includegraphics[width=10cm]{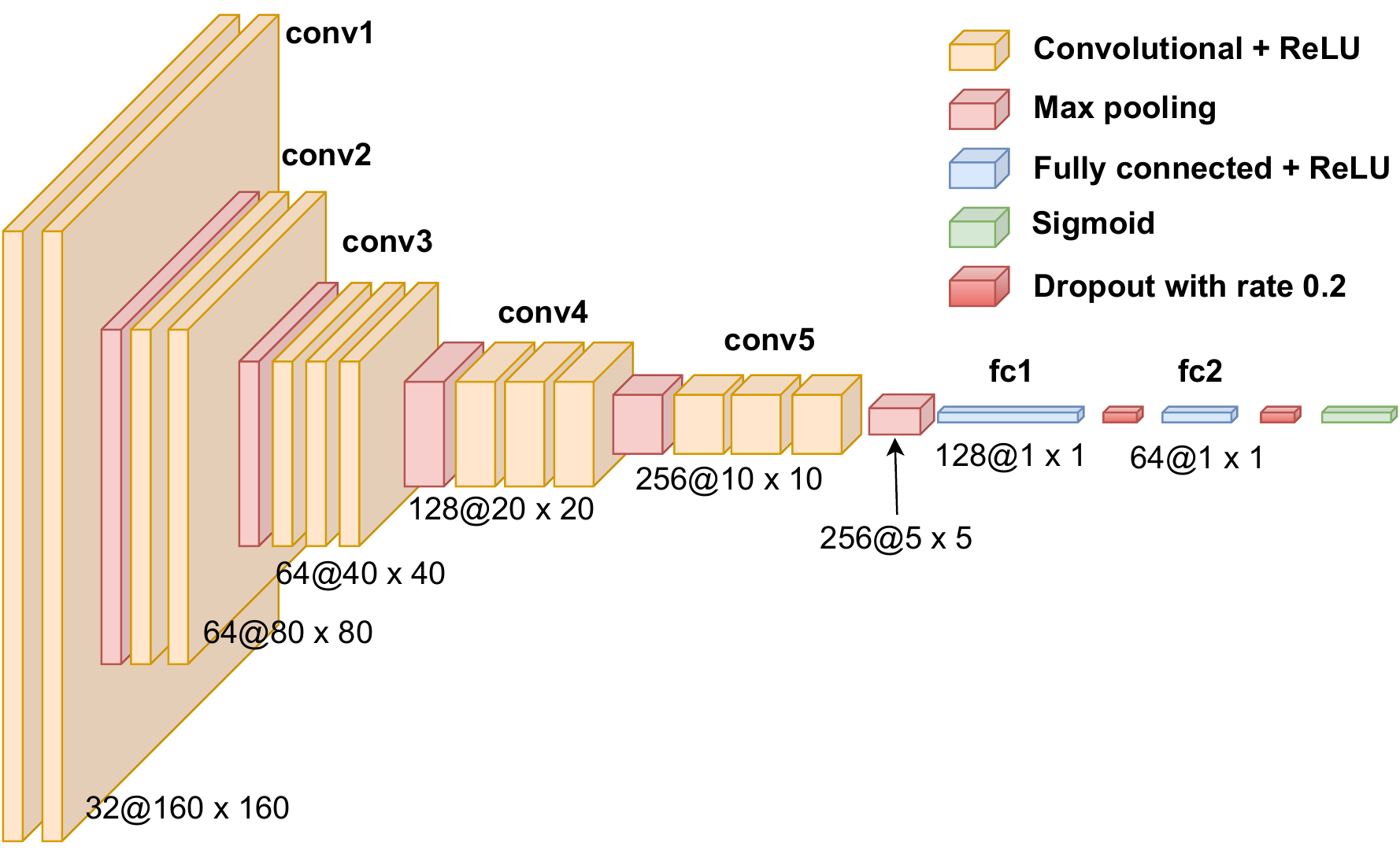}
   \caption{An illustration of the architecture of the anomaly detector, which is a modified version of the VGG16 network.}
  \label{fig:CNN_arch}
\end{figure}

A modified version of the VGG16 network was used as the anomaly detector classifier. The following modifications were made to the original network structure: the input size was decreased to 160 $\times$ 160, the number of filters in the hidden layers was decreased, dropout layers were added in between the fully connected layers, and the final softmax layer was replaced with a sigmoid layer. A normalising pre-processing layer was used to scale features between zero and one.
The resulting CNN with 23 layers has 2,847,777 trainable parameters.
The architecture is illustrated in figure \ref{fig:CNN_arch}, and a summary of the training data and parameters for the anomaly detector are given in table \ref{tab:CNN_training_ad}. In total, $\sim$10$^{6}$ training patches were used.

During inference, the expected normal-to-anomalous patch ratio is approximately 1,900. Such an imbalanced data set can lead to poor performance of the classifier, therefore, a data augmentation techniques and under-sampling the normal class were used on the training data set to bring this ratio to around 100. 
The anomalous patches were augmented by applying random and uniform brightness change in the range 0.75 - 1.25, multiples of 90-degree rotations, and horizontal and vertical flipping. In addition, the focal loss \cite{focal}, a dynamically weighted binary cross-entropy loss commonly used with imbalanced training data sets, was used as the loss function with default parameters $\gamma$ = 2 and $\alpha$ = 0.25.

\begin{table}[ht]
\centering
\caption{Summary of training data and parameters for the anomaly detector. Class 0 refers to normal and Class 1 to anomalous.}
\label{tab:CNN_training_ad}
\vspace{10pt}
\begin{tabular}{@{}ll@{}}
\toprule
Whole images       & 2,813   \\ 
Class 0 training patches    & 902,496 \\ 
Class 1 training patches    & 1,465    \\
\hspace{26pt}after augmentation  & 8,790  \\ 
Batch size      & 256 \\ 
Epochs            &  20 \\ 
Optimizer       & $Adam$ \\
Learning rate     &  10$^{-4}$  \\ 
Loss           & Focal loss\\ \bottomrule
\end{tabular}
\end{table}

\subsection{Validation}

In production, the pre-selected whole images are shown to an inspector. In addition, a fraction (10\%) of normal images are added to the set shown to the inspector to ensure the minimization of false negatives. The inspector either accepts, rejects or adds anomalous patches to validate the predictions. If only pre-selected images would have been shown, there would be a natural tendency to approve all images as anomalous and trust the PSA.
The validated data set is used as ground truths for performance monitoring and for continuous learning to incrementally improve the accuracy of the PSA.

\section{Results}
\label{sec:results}

\begin{figure}[h!]
\centering
  \includegraphics[width=6cm]{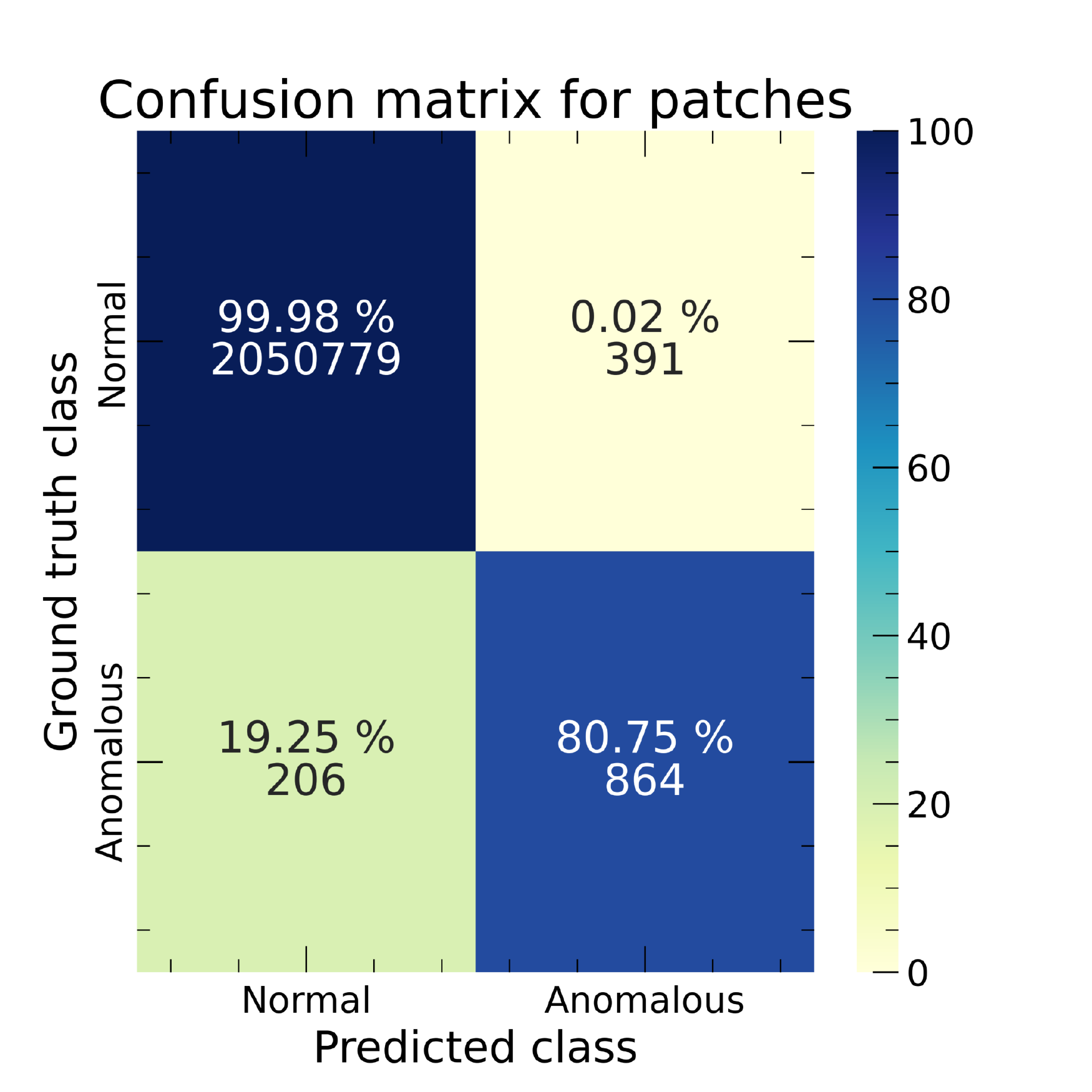}
  \includegraphics[width=6cm]{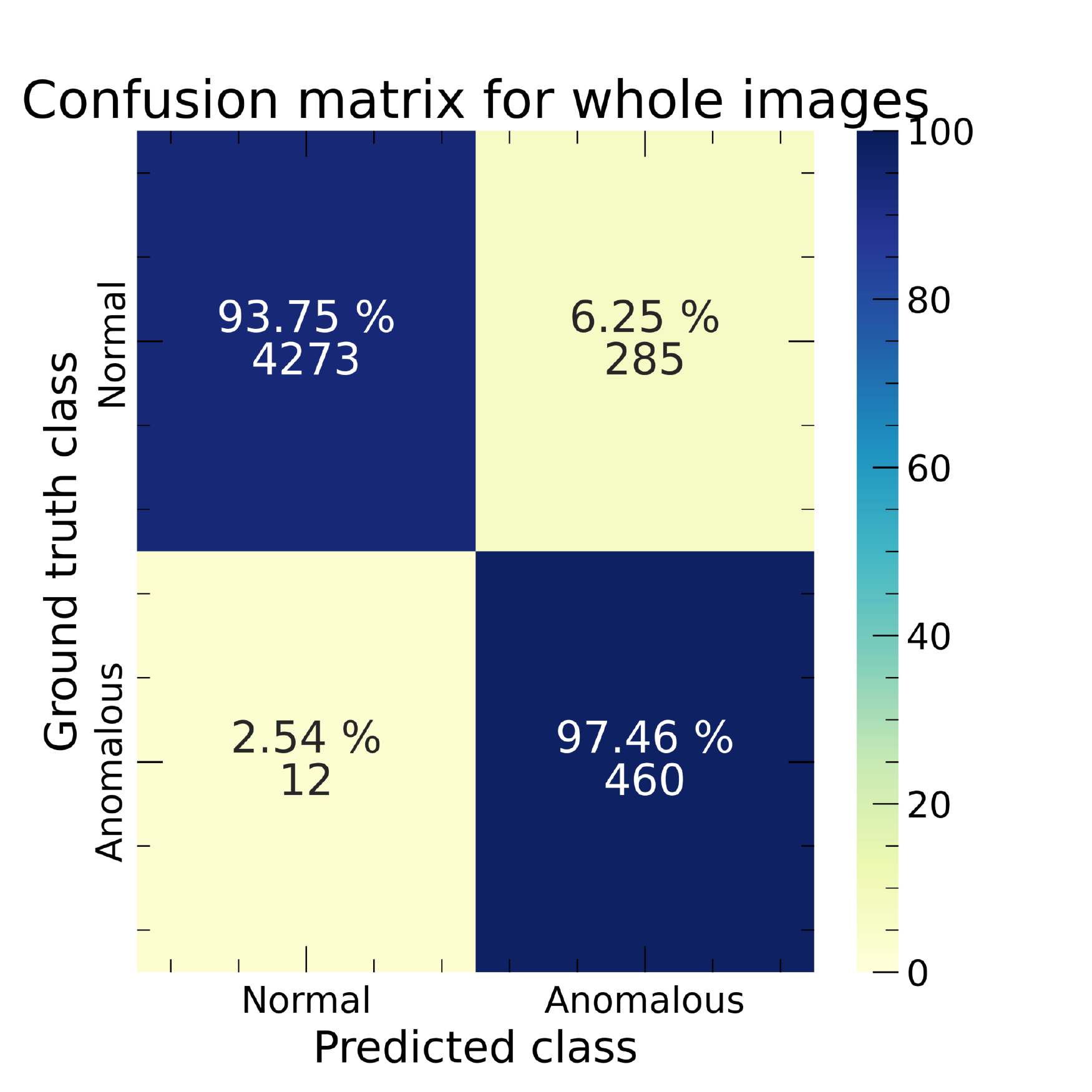}
  \caption{Confusion matrices for the patches (left) and the whole images (right). Top row corresponds to the confusion matrix normalized over ground truths.}
\label{fig:cms}
\end{figure}

Fifteen sensor scans acquired after the deployment of the PSA, corresponding to 2,052,240 patches from 5,030 whole images, were manually given ground truth labels.
The absolute and normalized confusion matrices for the patches are shown in figure \ref{fig:cms} (left). 
Due to data imbalance, the classification threshold must be set so that FPR for patches is significantly lower than the FNR to achieve an acceptable FPR for the whole images.

\begin{figure}[ht]
\centering
  \includegraphics[width=12cm]{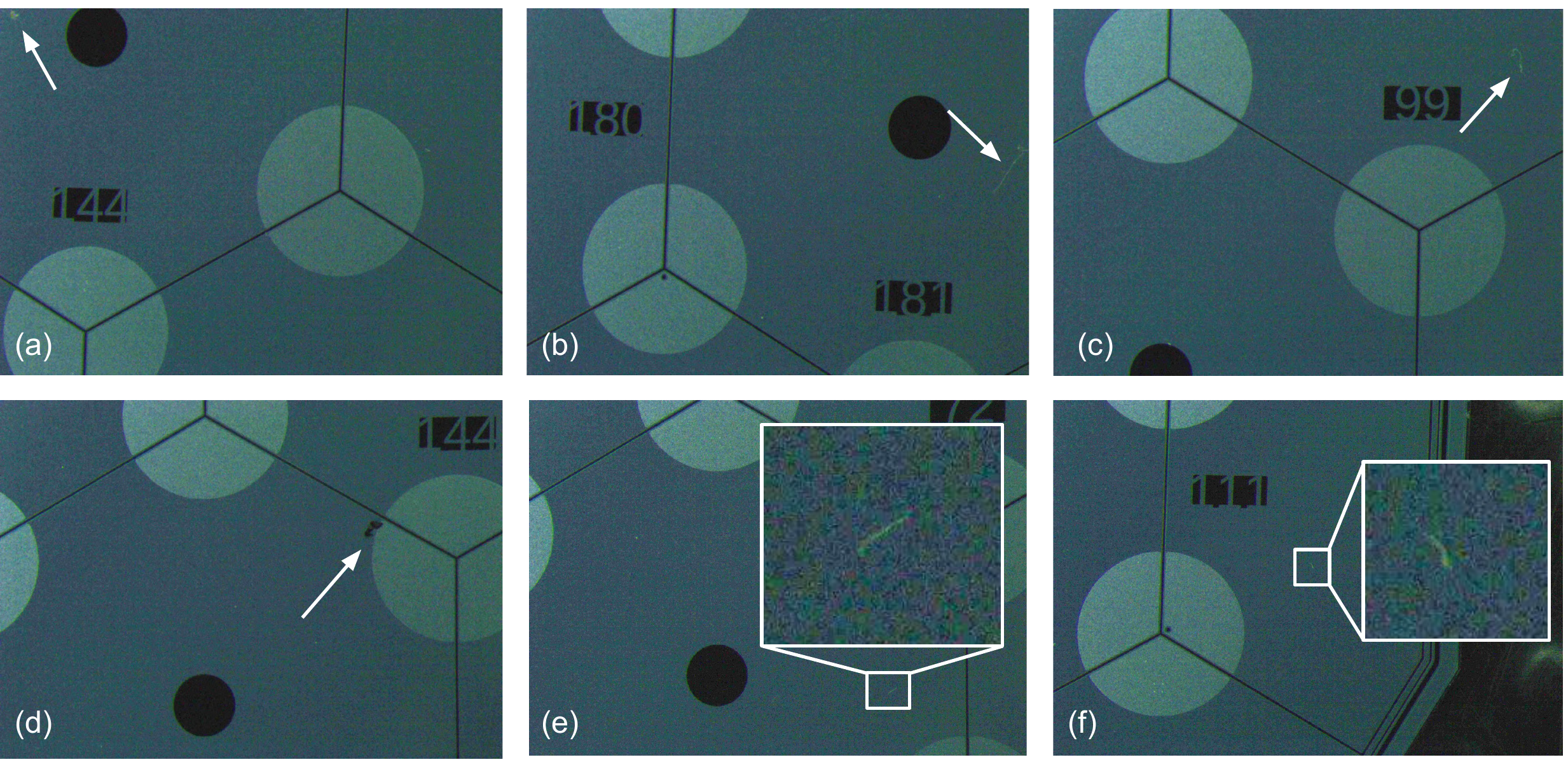}
  \caption{Examples of missed anomalous whole images. (a) A light stain (b, c) light dust particle (d) novel black anomaly (e, f) minor dust particle.}
  \label{fig:fns}
\end{figure}

A more relevant evaluation metric for the PSA is its performance on whole images.
A whole image is pre-selected if one or more patches are classified as anomalous.
The confusion matrices for the whole images are shown in figure \ref{fig:cms} (right), and evaluation metrics are reported in table \ref{tab:results}. 
Twelve anomalous whole images are incorrectly not pre-selected, resulting in an FNR of 2.5\%. 
Two missed images were of novel anomalies and ten were of light scratches and small dust particles.
Examples of missed anomalous whole images are shown in figure \ref{fig:fns}.

\begin{table}[ht]
\centering
\caption{Results for 5,030 whole images. Metrics as defined in \cite{metrics}.}
\label{tab:results}
\vspace{10pt}
\begin{tabular}{@{}ll@{}}
\toprule
Metric            & Value [\%] \\ \midrule
Recall            & 97.46  \\
Specificity       & 93.75  \\
Precision         & 61.75  \\
False negative rate       & \lineup\02.54   \\
False positive rate  & \lineup\06.25  \\
F-score                & 75.60  \\
Balanced accuracy & 95.60 \\\bottomrule
\end{tabular}
\end{table}

\begin{figure}[h!]
\centering
  \includegraphics[width=12cm]{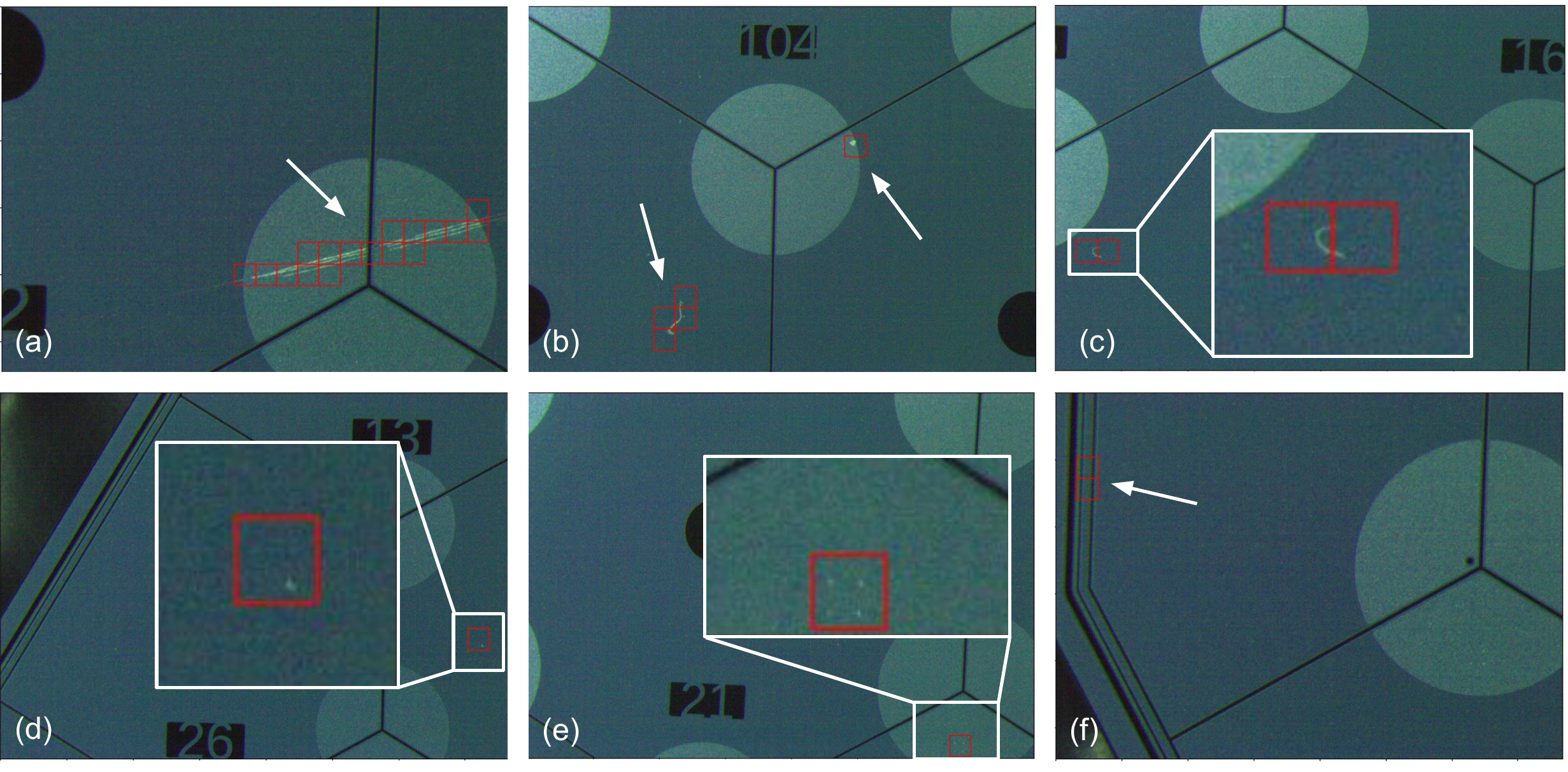}
  \caption{Examples of whole images pre-selected with the annotated patches. (a) A large scratch (b) dust particle and a stain (c) small dust particle (d) false positive (e) false positive on contact marks (f) false positive on a guard ring.}
  \label{fig:ps}
\end{figure}

The FNR of whole images is lower compared to the FNR of patches because most anomalous whole images have multiple anomalous patches. 
Thus, as long as at least one of the anomalous patches is classified correctly, the whole image is selected.
The FPR of whole images is less than 10\%. Examples of whole images pre-selected to be anomalous are shown in figure \ref{fig:ps}, where images (d-f) are false positives.
In total, 85\% of all whole images can be considered normal and do not require any human inspection.

\section{Discussion}
\label{sec:discussion}

\subsection{False negatives and positives}

It was observed that false negatives can sometimes be attributed to the fixed grid.
If only the default grid is used during inference, anomalies overlapping with or close to the grid lines can be missed. 
A proposed method for combating this is applying a secondary grid, which is illustrated in figure \ref{fig:inference_pipe}. 
The secondary grid is 16 $\times$ 23 patches in size, and it is overlaid on top of the default grid so that the patching is shifted from the top left corner by 80 pixels in both directions.
Only the whole images not selected using the default grid would then be evaluated with the additional secondary grid.

The anomaly detector tends to select patches at a guard ring\footnote{A guard ring is a structure at the periphery of a sensor designed to protect it from currents from the cutting edge, and shown in figure \ref{fig:ps} (f) as the black line with false positives.}, cell numbers and cell borders as false positive, due to the increase in the AE reconstruction error in these areas.
In addition, contact marks are a common source of false positives, see (e) in figure \ref{fig:ps}.
With the proposed approach, where the detection is invariant to the anomaly location on the sensor, these cannot be eliminated. 

\subsection{Efficiency}

Average inference times for a whole image are reported in table \ref{tab:inference_times}, with both the default grid and the secondary grid.
An average sensor scan consisted of 335 images, for which the picture-taking time is 9.2 minutes. 
On average, an inspector inspects a whole image in two seconds, evaluating one scan in 11 minutes.
On an RTX A2000, the inference time is less than 6 minutes for the scan with both grids. 
The GPU is necessary, as the short run time allows parallel picture-taking and pre-selection. At variance, the evaluation of a scan would take approximately 45 minutes on a regular CPU, significantly delaying the subsequent electrical characterization.

\begin{table}[ht]
\centering
\caption{Average inference times on a GPU or a CPU for whole images using the default and secondary grids.}
\label{tab:inference_times}
\vspace{10pt}
\begin{tabular}{@{}lllll@{}}
\toprule
\multicolumn{2}{l}{}          & Grid  &  Run time [s]     \\ \midrule
\multirow{3}{*}{} & RTX A2000 GPU & Default   & \lineup\00.8  \\
                           &     & + secondary & \lineup\01.0      \\
                           & CPU & Default   & \lineup\08.1    \\
                           &     & + secondary & 13.8         \\ \bottomrule
\end{tabular}
\end{table}

\subsection{Continuous learning}

Given that new data will be measured continuously, the PSA should be retrained to adapt to the new data.
Using new normal images to train the AE would not improve its task of poor anomaly reconstruction, so it does not require retraining.
Also the background detector is considered to be robust enough to not require immediate retraining.
However, the accuracy of the anomaly detector could be improved via training with novel anomalous patches.
For example, training instances such as the missed anomaly (d) in figure \ref{fig:fns} did not exist in the original training data set.

Anomalous images acquired after the initial deployment were given ground truth labels to extend the training data set.
The anomaly detector was retrained starting from randomly initialized weights with 2.1 times more anomalous patches than the original model.
An independent test data set was used to compare the original and retrained models.
The test data consists of 136 anomalous whole images and 754 normal whole images, for which the test metrics are presented in table \ref{tab:retrain}.
The retrained model performs significantly better on the test set.

\begin{table}[ht]
\centering
\caption{The performance of the pre-selection algorithm using the original and retrained anomaly detectors. Metrics are calculated for whole images using an independent test set.}
\label{tab:retrain}
\vspace{10pt}
\begin{tabular}{@{}lll@{}}
\toprule
Metric            &  Value [\%]   &        \\ 
                  & Original & Retrained  \\ \midrule
Recall            & 94.9      & \textbf{96.3} \\
Specificity       & 85.0      & \textbf{87.8} \\
Precision         & 53.3     & \textbf{58.7} \\
False negative rate        & \lineup\05.2      & \hspace{0.6mm}\lineup\0\textbf{3.7} \\
False positive rate  & 15.0    & \textbf{12.2} \\ 
F1-score  & 68.0    & \textbf{72.9} \\ 
Balanced accuracy & 90.0    & \textbf{92.1} \\ \bottomrule
\end{tabular}
\end{table}

\section{Conclusion} 
\label{sec:conclusion}
A deep learning-based pre-selection algorithm (PSA) that fully automates the visual inspection (VI) of the silicon sensors produced for the construction of the CMS HGCAL detector was developed. An ensemble of a deep convolutional autoencoder and a neural network for classification is used, with a patching applied before the classification to allow the general localization of the anomalies in the images of the sensors. The automated VI is a vital part of the quality control (QC) of the analyzed sensors.

The performance of the PSA was evaluated using fifteen full sensor scans acquired in production in a clean room dedicated to sensor testing at CERN. 
The recall, which measures the fraction of anomalous images that are found, was 97.46\%, with an acceptable FPR of 6.25\%.
The images are evaluated in real-time, and approximately 85\% of all images can be discarded as normal, thus removing the need for human labor to inspect them.
The developed automated VI is standardized, and therefore also believed to be less biased by the subjectivity of a human inspector. On average, it saves 10 minutes of resources per sensor, and for each batch of hundred sensors, this corresponds to 17 person-hours less required to manually inspect the images. 

The accuracy is considered sufficient for deployment, even though the PSA was shown to fail to select a small fraction of images with light and small scratches and dust particles, in addition to novel types of anomalies. It was demonstrated that as more anomalous images are acquired in production, the data can be used to retrain the anomaly detection model to further improve the accuracy of the PSA.

A major advantage of the presented approach is its intrinsic generality. The algorithm acts on microscope images of the sensor surface, each of which covers only a small fraction of the total surface area of a full 8" sensor, and the anomaly detection is invariant to the image location on the sensor. Thus, the PSA is applicable to variable or incomplete scans, or partial sensors. 

Thanks to its generality, accuracy and speed, the presented architecture of the pre-selection and annotation model could be used in other applications of automating the detection of small anomalies from images taken in a changing environment.

%\ack{}
\ack{M.~P. and S.~G. are supported by the European Research Council (ERC) under the European Union's Horizon 2020 research and innovation program (grant agreement n$^o$ 772369).
The authors would like to thank Dr. Thorben Quast for guidance and suggestions.
The authors would also like to acknowledge the members of the CMS HGCAL group for their contributions to the collection of the data set.}

\section*{References}
\bibliographystyle{unsrt} 
\bibliography{references}

\end{document}